# Closed-form expressions for correlated density matrices: application to dispersive interactions and example of (He)$_2$

## Sébastien RAGOT[a)] and Pierre J. BECKER

Laboratoire Structure, Propriété et Modélisation des Solides (CNRS, Unité Mixte de Recherche 85-80). École Centrale Paris, Grande Voie des Vignes, 92295 CHATENAY-MALABRY, FRANCE

## Abstract

Empirically correlated density matrices of $N$-electron systems are investigated. Exact closed-form expressions are derived for the one- and two-electron reduced density matrices from a general pairwise correlated wave function. Approximate expressions are proposed which reflect dispersive interactions between closed-shell centro-symmetric subsystems. Said expressions clearly illustrate the consequences of second-order correlation effects on the reduced density matrices. Application is made to a simple example: the (He)$_2$ system. Reduced density matrices are explicitly calculated, correct to second order in correlation, and compared with approximations of independent electrons and independent electron pairs. The models proposed allow for variational calculations of interaction energies and equilibrium distance as well as a clear interpretation of dispersive effects on electron distributions. Both exchange and second order correlation effects are shown to play a critical role on the quality of the results.

[a)] at_home@club-internet.fr

# I. Introduction

Van der Waals (VDW) forces are an important class of attractive intermolecular forces which involve polarization of molecules. They notably include the noted London dispersion forces, which arise from temporarily induced dipoles and can therefore be exhibited by nonpolar atoms or molecules. Such interactions involve correlation of distant electrons. Dispersion interactions can accordingly not be accounted for at Hartree-Fock (HF) level.

Moreover, in the context of density functional theory (DFT), it has often been outlined[1,2] that usual (approximate) exchange-correlation functionals can neither reproduce realistic interaction energies nor lead to satisfactory equilibrium distances. For example, semilocal generalized gradient approximations may fail to predict any attraction between two spherically symmetric non-overlapping electron densities[3,4] while local density approximation (LDA) is rather overbinding.[5] In facts, the use of exchange-correlation functionals merely results in underestimated equilibrium distances and overestimated interaction energies.[1,2,3] In particular, it has been shown that the choice of the exchange functional is critical[1] for an accurate description of VDW systems, especially in the "bond" region. Therefore, modeling dispersion effects requires a careful analysis. To this aim, various approaches have been proposed, see for instance Refs. 1,4,5,6,7,8,9,10,11 and, in particular Refs. 10,11 for a discussion thereof.

Beyond the computation of interaction energies and intermolecular potentials, an issue of this paper is the modeling of electron distributions. In particular, questions that raise are: how (much) are electron densities impacted by dispersion effects? Can we possibly "see" dispersion effects through x-Ray experiments? In this respect, we adopt here a density-matrix description of electron correlation, as density matrices provide a natural basis for passing from direct to momentum space and vice versa.[12,13,14] Furthermore, electron correlation can be accounted for in the one-electron matrix and subsequently in the momentum density, in contrast with a DFT approach.[15]

In the following, closed-form expressions for the one- and two-electron reduced density matrices for a general $N$-electron system are derived from a pairwise correlated wavefunction. The approach chosen extends the original Colle-Salvetti's scheme to account for second-order correlation effects on both pair density and one-electron density-matrix (sect. II). In the case of dispersive correlation, simple correlation functions may be inferred from perturbation theory. Closed form expressions are then derived for electron distributions (sect. III). In the last section and as a check, use is made of said expressions for the computation of electronic properties of the $(He)_2$ system. Equilibrium distance and interaction energy are computed. The model density-matrices are compared to other approaches, including independent electron- and independent pair approximations.

## II. Empirically correlated density matrices

The spinless one- and two-electron reduced density-matrices (hereafter 1- and 2-RDMs) derived from a general $N$-electron wave function $\psi$ are defined as[16,17]

$$\rho_1(\mathbf{r}_1;\mathbf{r}_1') = N Tr_{s_1,x_2,\ldots,x_N}\left[\psi(\mathbf{r}_1 s_1, x_2,\ldots, x_N)\psi^*(\mathbf{r}_1' s_1, x_2,\ldots, x_N)\right], \tag{1}$$

and

$$\rho_2(\mathbf{r}_1,\mathbf{r}_2;\mathbf{r}_1',\mathbf{r}_2') = \frac{N(N-1)}{2} N Tr_{s_1,s_2,x_3,\ldots,x_N}\left[\psi(\mathbf{r}_1 s_1, \mathbf{r}_2 s_2,\ldots, x_N)\psi^*(\mathbf{r}_1' s_1, \mathbf{r}_2' s_2,\ldots, x_N)\right], \tag{2}$$

where $\mathbf{r}_i$, $s_i$ and $x_i$ are space, spin and global space-spin coordinates of electron $i$, respectively. From these definitions follows the condition

$$\rho_1(\mathbf{r}_1;\mathbf{r}_1') = \frac{2}{(N-1)} Tr_{\mathbf{r}_2}\left[\rho_2(\mathbf{r}_1,\mathbf{r}_2;\mathbf{r}_1',\mathbf{r}_2)\right]. \tag{3}$$

The diagonal elements of the 1- and 2-RDMs relate to (indirectly) observable quantities, which are the electron charge and pair densities, that is

$$\rho(\mathbf{r}_1) = \rho_1(\mathbf{r}_1; \mathbf{r}_1),$$

(4)

and

$$P(\mathbf{r}_1, \mathbf{r}_2) = \rho_2(\mathbf{r}_1, \mathbf{r}_2; \mathbf{r}_1, \mathbf{r}_2).$$

(5)

We have thus defined a set of distributions involved in the calculation of the exact non-relativistic energy $E$ in the Born-Oppenheimer approximation.[17]

A classic trial correlated wave function (WF) is[18]

$$\psi = \psi^{(0)}(x_1, x_2, ..., x_N) \prod_{i<j} (1 + \omega_{ij}),$$

(6)

where $\omega_{ij}$ (assumed real) correlates in space the electron pair $(i, j)$ and $\psi^{(0)}$ denotes a non-correlated approximation to the exact ground-state WF, e.g. a single determinant. From Eq. (6), the 2-RDM can be developed as[19]

$$\rho_2(\mathbf{r}_1, \mathbf{r}_2; \mathbf{r}_1', \mathbf{r}_2') = \rho_2^{(0)}(\mathbf{r}_1, \mathbf{r}_2; \mathbf{r}_1', \mathbf{r}_2')(1 + \omega_{12} + \omega_{1'2'} + \omega_{12}\omega_{1'2'}) + R(\mathbf{r}_1, \mathbf{r}_2; \mathbf{r}_1', \mathbf{r}_2'),$$

(7)

where $\rho_2^{(0)}$ is the determinantal 2-RDM and $R$ includes all the terms that can not be factorized as $\rho_2^{(0)}$ times a correlation factor, i.e. integrals involving $\rho_3^{(0)}$ and higher-order RDMs. Neglecting $R$ yields

$$\rho_2(\mathbf{r}_1, \mathbf{r}_2; \mathbf{r}_1', \mathbf{r}_2') = \rho_2^{(0)}(\mathbf{r}_1, \mathbf{r}_2; \mathbf{r}_1', \mathbf{r}_2')(1 + \omega_{12} + \omega_{1'2'} + \omega_{12}\omega_{1'2'}).$$

(8)

The approach of Colle-Salvetti[20] (hereafter CS) makes Eq. (8) the starting point for the derivation of a correlation energy expression. This expression bypasses $N$-electron effects (the neglected $R$ term) on the pair density beyond those involving one pair at a time. As such, this approach has some connection with an independent pair approximation (IPA), which is known to be correct to first order only in correlation[21]. We note, however, that in spite of such physical inconsistencies[15,22,23] the CS scheme is particularly simple and surprisingly accurate.[24,25,26]

As discussed in Ref. 26, the correlation function $\omega_{ij}$ is empirically parameterized in the CS scheme. Accordingly, calculation of the correlation energy is not variational. Note that even if $\omega_{ij}$ were variationaly optimized, it could manifestly not lead to correct energies, since (i) the model 2-RDM of Eq. (8) is not correctly normalized and (ii) the model assumes somehow independent pairs. Further, calculating the 1-RDM from Eq. (8), using Eq. (3), results in discarding most of the correlation effects on one-electron densities and thus the kinetic counterpart of correlation, which must at least partly balance the correlation-induced lessening of potential energy.[26] This point shall be exemplified below.

In order to get exact closed-form expressions for the RDMs, we can write the trial wavefunction as

$$\psi = \mathcal{N}^{1/2}\left(1 + \sum_{i<j}\omega_{ij}\right)\psi^{(0)},$$ (9)

leading to

$$\rho_2 = \mathcal{N}\frac{N(N-1)}{2}Tr_{s_1,s_2,3\ldots N}\left[\psi^{(0)}\psi^{(0)*}\left\{1 + \sum_{i<j}\left(\omega_{ij} + \omega_{i'j'} + \sum_{k'<l'}\omega_{ij}\omega_{k'l'}\right)\right\}\right]$$ (10)

and

$$\rho_1 = \mathcal{N}N\,Tr_{s_1,2\ldots N}\left[\psi^{(0)}\psi^{(0)*}\left\{1 + \sum_{i<j}\left(\omega_{ij} + \omega_{i'j'} + \sum_{k'<l'}\omega_{ij}\omega_{k'l'}\right)\right\}\right],$$ (11)

where some variables are omitted for clarity. From Eqs. (10) and (11), exact closed-form expressions can be derived for the 1- and 2-RDMs[26], involving up to the 5- and 6-electron uncorrelated matrices, respectively.

More tractable expressions may for instance be obtained by constraining the functions $\omega_{ij}$ to satisfy the rules

$$Tr_j\left[\psi^{(0)}(x_1,x_2,...,x_N)\psi^{(0)*}(x_1',x_2,...,x_N)\{\omega_{ij}\}\right]=0 \qquad (i\neq j)$$

$$Tr_l\left[\psi^{(0)}(x_1,x_2,...,x_N)\psi^{(0)*}(x_1',x_2,...,x_N)\{\omega_{ij}\omega_{kl}\}\right]=0 \qquad (l\neq\{k,i,j\})$$
(12)

Such conditions are readily satisfied by writing $\omega$'s as a suitable sum of operators coupling occupied orbitals to virtual orthogonal orbitals. Then, the resulting RDMs reduce to

$$\rho_1=\mathcal{N}\left\{\rho_1^{(0)}+2Tr_2\left[\left(\omega_{12}\omega_{1'2}^*\right)\rho_2^{(0)}\right]+3Tr_{23}\left[\left(\omega_{23}\omega_{23}^*\right)\rho_3^{(0)}\right]\right\},$$
(13)

and

$$\rho_2=\mathcal{N}\left\{\rho_2^{(0)}\left(1+\omega_{12}+\omega_{1'2'}^*+\omega_{12}\omega_{1'2'}^*\right)\right.$$
$$\left.+3Tr_3\left[\left(\omega_{13}+\omega_{23}\right)\left(\omega_{1'3}^*+\omega_{2'3}\right)\rho_3^{(0)}\right]+6Tr_{34}\left[\omega_{34}\omega_{34}^*\rho_4^{(0)}\right]\right\}$$
(14)

Note that Eq. (14) differs substantially from Eq. (8) through terms or order $\omega^2$. (He)re, Eqs. (13) and (14) satisfy the condition defined by Eq. (3).

## III.  Dispersive interactions

Let us now consider two neutral closed-shell subsystems $a$ and $b$, interacting at typical VDW distances. The Hamiltonian decomposes as $H=H_A+H_B+W_{AB}$ where $W_{AB}$ includes all intermolecular potential terms:

$$W_{AB}=\frac{Z_AZ_B}{R_{AB}}-\sum_{i\in A}\frac{Z_B}{r_{iB}}-\sum_{j\in B}\frac{Z_A}{r_{jA}}+\sum_{\substack{i\in A\\j\in B}}\frac{1}{r_{ij}}.$$

First, in an independent electron (IE) approach, the determinantal pair density can be written as[17]

$$P_{IE}^{(0)}(\mathbf{r}_1,\mathbf{r}_2)=\frac{1}{2}\left\{\rho_{1,IE}^{(0)}(\mathbf{r}_1;\mathbf{r}_1)\rho_{1,IE}^{(0)}(\mathbf{r}_2;\mathbf{r}_2)-\frac{1}{2}\rho_{1,IE}^{(0)}(\mathbf{r}_1;\mathbf{r}_2)\rho_{1,IE}^{(0)}(\mathbf{r}_2;\mathbf{r}_1)\right\}.$$
(15)

where $\rho_{1,IE}^{(0)}(\mathbf{r}_1;\mathbf{r}_1)$ is obtained from a determinantal wavefunction through Eq. (1). This expression takes care of Pauli repulsion but does obviously not reflect attractive dispersion interactions.

Next, in order to go beyond the IE approach and find a suitable trial wavefunction, we shall consider the perturbation theory. As known, the first-order perturbated ground-state $\psi_0^{(1)}$ of a system experiencing a perturbation $W$ writes as

$$\psi_0^{(1)} = \psi_0^{(0)} + \sum_{n \neq 0} \frac{\left\langle \psi_n^{(0)} \left| \hat{W} \right| \psi_0^{(0)} \right\rangle}{\Delta_{0n}} \psi_n^{(0)}, \tag{16}$$

where $\Delta_{0n} = E_0 - E_n$ is the difference between energies corresponding to states $\psi_0^{(0)}$ and $\psi_n^{(0)}$. For neutral subsystems, $W$ can be developed as:

$$W = \sum_{i \in a, \; j \in b} \frac{1}{R_{ab}^3} \left( x_{ia} x_{jb} + y_{ia} y_{jb} - 2 z_{ia} z_{jb} \right) + \ldots,$$

where $x_{ic}$ is the Cartesian displacement of electron $i$ from the centroid of negative charge of subsystems c, which are located along the z-axis. In the following, each of the subsystems will be assumed centro-symmetric, for simplicity.

Replacing now all excitation energies $\Delta_{0n}$ by an average excitation energy $\Delta$ (i.e. the Unsöld approximation[27]) and using the closure relation, we get the following approximation of $\psi_0^{(1)}$

$$\psi_0^{(1)} = \left\{ 1 + \frac{1}{\Delta} \left( W - \left\langle \psi_0^{(0)} \left| \hat{W} \right| \psi_0^{(0)} \right\rangle \right) \right\} \psi_0^{(0)} \tag{17}$$

The parameter $\Delta$ may further be considered as a variational parameter.

A trial ground-state wavefunction can thus be rewritten

$$\psi = \psi^{(0)}(x_1, x_2, \ldots, x_N) \left\{ 1 + \sum_{i \in a, \; j \in b} \left( \omega(\mathbf{r}_{ia}, \mathbf{r}_{jb}) - \left\langle \psi^{(0)} \left| \omega(\mathbf{r}_{ia}, \mathbf{r}_{jb}) \right| \psi^{(0)} \right\rangle \right) \right\}, \tag{18}$$

where $\omega(\mathbf{r}_{ia}, \mathbf{r}_{jb}) = \frac{1}{\Delta R_{ab}^3} \left( x_{ia} x_{jb} + y_{ia} y_{jb} - 2 z_{ia} z_{jb} \right)$ correlates two distant electrons $(i, j)$, respectively "located" near subsystems $a$ and $b$.

In order to carry out explicit calculations for the RDMs, we introduce partial RDMs $\rho_{1,c}^{(0)}(\mathbf{r}_1,\mathbf{r}_{1'})$ and $\rho_{2,c}^{(0)}(\mathbf{r}_1,\mathbf{r}_2;\mathbf{r}_{1'},\mathbf{r}_{2'})$, with $c = a$ or $b$. Their diagonal part $\rho_c^{(0)}(\mathbf{r}_1) = \rho_{1,c}^{(0)}(\mathbf{r}_1,\mathbf{r}_1)$ and $P_c^{(0)}(\mathbf{r}_1,\mathbf{r}_2) = \rho_{2,c}^{(0)}(\mathbf{r}_1,\mathbf{r}_2;\mathbf{r}_1,\mathbf{r}_2)$ integrate to $N_c$ and $N_c$ ($N_c$ - 1)/2, respectively.

Next, the interaction energy of nonpolar and weakly polarizable systems is known to be dominated by the first three terms in the usual expansion:

$$E_{\text{int}} = E_{pol}^{(1)} + E_{exch}^{(1)} + E_{pol}^{(2)} + ...,$$

where $E_{pol}^{(1)}$ is the damped classical electrostatic interaction energy, $E_{pol}^{(2)} = E_{ind}^{(2)} + E_{disp}^{(2)}$ is the sum of the damped classical induction and dispersion energies and $E_{exch}^{(1)}$ is the first exchange correction, as defined by symmetry-adapted perturbation theories (SAPT)[28].

This suggests neglecting the coupling between exchange and correlation in our approach. For instance, using $\omega(\mathbf{r}_{ia},\mathbf{r}_{jb}) = \dfrac{1}{\Delta R_{ab}^3}\left(x_{ia}x_{jb} + y_{ia}y_{jb} - 2z_{ia}z_{jb}\right)$ leads to the following approximation

$$
\begin{aligned}
\left\langle\psi^{(0)}\left|\omega(\mathbf{r}_{ia},\mathbf{r}_{jb})\right|\psi^{(0)}\right\rangle &\approx \left\langle\psi_a^{(0)}\psi_b^{(0)}\left|\omega(\mathbf{r}_{ia},\mathbf{r}_{jb})\right|\psi_a^{(0)}\psi_b^{(0)}\right\rangle = 0 \\
\left\langle\psi^{(0)}\left|\omega(\mathbf{r}_{ia},\mathbf{r}_{jb})\omega(\mathbf{r}_{ka},\mathbf{r}_{lb})\right|\psi^{(0)}\right\rangle &\approx \left\langle\psi_a^{(0)}\psi_b^{(0)}\left|\omega(\mathbf{r}_{ia},\mathbf{r}_{jb})\omega(\mathbf{r}_{ka},\mathbf{r}_{lb})\right|\psi_a^{(0)}\psi_b^{(0)}\right\rangle = 0 \left(l \neq \{k,i,j\}\right),
\end{aligned}
\tag{19}
$$

since centrosymmetric subsystems have no net dipoles.

Bearing in mind the above approximations, we arrive at the following expressions for the model "dispersive" RDMs:

$$
\begin{aligned}
\rho_1(\mathbf{r}_1;\mathbf{r}_1')/\mathcal{N} &= \rho_{1,IE}^{(0)}(\mathbf{r}_1;\mathbf{r}_1') \\
&+ (1+\mathscr{P}_{ab})\rho_{1,a}^{(0)}(\mathbf{r}_1;\mathbf{r}_1')\left(N_b\left\langle\omega(\mathbf{r}_{1a},\mathbf{r}_{2b})\omega(\mathbf{r}_{1a}',\mathbf{r}_{2b})\right\rangle_2 + (N_a-1)N_b\left\langle\omega(\mathbf{r}_{2a},\mathbf{r}_{3b})^2\right\rangle_{2,3}\right),
\end{aligned}
\tag{20}
$$

and

$$
\begin{aligned}
P_2(\mathbf{r}_1,\mathbf{r}_2)/\mathscr{N} =\ & \\
& P_{IE}^{(0)}(\mathbf{r}_1,\mathbf{r}_2) \\
& + \tfrac{1}{2}\left(1+\mathscr{P}_{ab}\right)\rho_a^{(0)}(\mathbf{r}_{1a})\rho_b^{(0)}(\mathbf{r}_{2b})\times \\
& \quad \left\{ 2\omega(\mathbf{r}_{1a},\mathbf{r}_{2b}) + \omega(\mathbf{r}_{1a},\mathbf{r}_{2b})^2 \right. \\
& \quad \left. + (N_b-1)\left\langle\omega(\mathbf{r}_{1a},\mathbf{r}_{3b})^2\right\rangle_3 + (N_a-1)\left\langle\omega(\mathbf{r}_{3a},\mathbf{r}_{2b})^2\right\rangle_3 + (N_a-1)(N_b-1)\left\langle\omega(\mathbf{r}_{3a},\mathbf{r}_{4b})^2\right\rangle_{3,4} \right\} \\
& + \left(1+\mathscr{P}_{ab}\right)P_a^{(0)}(\mathbf{r}_1,\mathbf{r}_2)\times \\
& \quad \left\{ N_b\left\langle\left(\omega(\mathbf{r}_{1a},\mathbf{r}_{3b})+\omega(\mathbf{r}_{2a},\mathbf{r}_{3b})\right)^2\right\rangle_3 + (N_a-2)N_b\left\langle\omega(\mathbf{r}_{3a},\mathbf{r}_{4b})^2\right\rangle_{3,4} \right\}
\end{aligned}
\tag{21}
$$

where $\mathscr{P}_{ab}$ permutes $a$ and $b$ indices. Note that $\rho(\mathbf{r}_1)=\rho_1(\mathbf{r}_1;\mathbf{r}_1)$ and $P_2(\mathbf{r}_1,\mathbf{r}_2)$ as defined in

Eqs. (20) and (21) are consistent in the sense of Eq. (3), upon integration over $\mathbf{r}_2$. Moreover,

Eq. (20) can be equivalently formulated in momentum space through the correspondence:

$n(\mathbf{p}_1,\mathbf{p}_{1'}) = \frac{1}{(2\pi)^3}\int \rho(\mathbf{r}_1,\mathbf{r}_{1'})e^{i(\mathbf{p}_1\cdot\mathbf{r}_1-\mathbf{p}_{1'}\cdot\mathbf{r}_{1'})}d\mathbf{r}_1 d\mathbf{r}_{1'}$, and $x \equiv \partial/\partial p_x$ .

Some comments are in order.

(i) Terms like $\omega^2$ encompass second-order correlation effects. In particular, only second-order correlation effects survive in the 1-RDM, Eq. (20).

(ii) The first term $P_{IE}^{(0)}$ in Eq. (21) is the determinantal pair density; it includes zero-order exchange between subsystems $a$ and $b$.

(iii) The term $\tfrac{1}{2}\left(1+\mathscr{P}_{ab}\right)\rho_a^{(0)}(\mathbf{r}_{1a})\rho_b^{(0)}(\mathbf{r}_{2b})$ in Eq. (21) has the meaning of an uncorrelated inter-subsystem pair density; $\omega(\mathbf{r}_{1a},\mathbf{r}_{2b})$ reflect first-order correlation between electrons 1 and 2 in the reference pair.

(iv) While $\omega(\mathbf{r}_{1A},\mathbf{r}_{2B})^2$ involves electrons 1 and 2 only, the average terms reflect effects of various pairs of distant electrons on electron 1 and 2. For example, $\left\langle\omega(\mathbf{r}_{3a},\mathbf{r}_{4b})^2\right\rangle_{3,4}$ bears effects on the reference pair (1,2) due to a distinct electron pair. Such terms carry $N$-electron effects induced by the pairwise correlated wavefunction, which are not accounted for in a CS-like approach.

Next, explicit expressions for the terms $\left\langle \omega(\mathbf{r}_{ia},\mathbf{r}_{kb})\omega(\mathbf{r}_{ja}',\mathbf{r}_{kb})^* \right\rangle_k$ can be obtained. To this aim, we introduce functions $f(\mathbf{r}_{ic},\mathbf{r}_{jc})$, defined as

$$f_b(\mathbf{r}_{1a},\mathbf{r}_{1'a}) = \frac{1}{\Delta^2 R_{ab}^6}\left(\sigma_{xb}x_{1a}x_{1'a} + \sigma_{yb}y_{1a}y_{1'a} + 4\sigma_{zb}z_{1a}z_{1'a}\right) \qquad (22)$$

$$f_a(\mathbf{r}_{1b},\mathbf{r}_{1'b}) = \frac{1}{\Delta^2 R_{ab}^6}\left(\sigma_{xa}x_{1b}x_{1'b} + \sigma_{ya}y_{1b}y_{1'b} + 4\sigma_{za}z_{1b}z_{1'b}\right)$$

$$\bar{f} = \frac{1}{\Delta^2 R_{ab}^6}\left(\sigma_{xa}\sigma_{xb} + \sigma_{ya}\sigma_{yb} + 4\sigma_{za}\sigma_{zb}\right) \qquad,$$

where $\sigma_{xc} = \frac{1}{N_c}\int x_{1c}^2 \rho_c^{(0)}(r_1,r_1)dr_1$ is the quadratic spatial extension per electron. In addition, the spatial extension further simplifies as $\sigma_{xc} = \sigma_{yc} = \sigma_{zc}$, due to the assumed symmetry of the sub-systems.

With these notations and neglecting the coupling of exchange and correlation between distant electrons, we have:

$$\left\langle \omega(\mathbf{r}_{ia},\mathbf{r}_{kb})\omega(\mathbf{r}_{ja}',\mathbf{r}_{kb}) \right\rangle_k = f_b(\mathbf{r}_{ia},\mathbf{r}_{ja}')$$

and

$$\left\langle \omega(\mathbf{r}_{ia},\mathbf{r}_{kb})^2 \right\rangle_{i,k} = \bar{f} .$$

Alternatively, functions $f(\mathbf{r}_{ic},\mathbf{r}_{jc})$ and $\bar{f}$ can further be expressed in terms of linear polarizabilities of sub-systems, using again the Unsöld approximation.[29]

Therefore, the model proposed is fully analytical. The parameters governing the magnitude of deformations of electron distributions are for instance the average excitation energies or the sub-systems linear polarizabilities and quadratic spatial extensions in each subsystems.

# IV. Application to the cohesive energy of (He)$_2$

We now consider two similar closed-shell systems of two electrons each. Let the two functions $a$ and $b$ correspond to different locations, with $\mathbf{a}$ and $\mathbf{b}$ pointing at the corresponding centers and $R_{ab}$ be the (supposedly large) distance between the two centers. In a valence-bond approach, the two occupied orbitals are $\varphi_+$ and $\varphi_-$, defined as

$$\varphi_\pm = \frac{1}{\sqrt{2(1 \pm S)}}(a \pm b),$$

so that in an independent electron (IE) approach, the corresponding spinless 1-RDM can be simply written as: $\rho_{1,IE}^{(0)}(\mathbf{r};\mathbf{r}') = 2\varphi_+(\mathbf{r})\varphi_+(\mathbf{r}') + 2\varphi_-(\mathbf{r})\varphi_-(\mathbf{r}')$. The corresponding pair density follows from Eq. (15) and the matrices of each subsystem are in this case

$$\rho_{1,c}^{(0)}(\mathbf{r}_1,\mathbf{r}_{1'}) \equiv 2c(\mathbf{r}_1)c(\mathbf{r}_{1'})$$

and

$$\rho_{2,c}^{(0)}(\mathbf{r}_1,\mathbf{r}_2;\mathbf{r}_{1'},\mathbf{r}_{2'}) = c(\mathbf{r}_1)c(\mathbf{r}_2)c(\mathbf{r}_{1'})c(\mathbf{r}_{2'}),$$

The densities (including dispersive effects) obtained from Eqs. (20) and (21) thus reduces to:

$$\rho_1(\mathbf{r}_1,\mathbf{r}_{1'}) = \mathcal{N}\left\{ \rho^{(0)}(\mathbf{r}_1,\mathbf{r}_{1'}) + 2\rho_a^{(0)}(\mathbf{r}_{1a},\mathbf{r}_{1a'})\left[ f(\mathbf{r}_{1a},\mathbf{r}_{1a'}) + \bar{f} \right] + 2\rho_b^{(0)}(\mathbf{r}_{1b},\mathbf{r}_{1b'})\left[ f(\mathbf{r}_{1b},\mathbf{r}_{1b'}) + \bar{f} \right] \right\} \quad (23)$$

and

$$
\begin{aligned}
P_2(\mathbf{r}_1,\mathbf{r}_2) = \mathcal{N}\Big\{ & P_{IE}^{(0)}(\mathbf{r}_1,\mathbf{r}_2) \\
& + (1 + \mathscr{P}_{ab})\tfrac{1}{2}\rho_a^{(0)}(\mathbf{r}_{1a})\rho_b^{(0)}(\mathbf{r}_{2b})\left\{ 2\omega(\mathbf{r}_{1a},\mathbf{r}_{2b}) + \omega(\mathbf{r}_{1a},\mathbf{r}_{2b})^2 + f(\mathbf{r}_{1a},\mathbf{r}_{1a}) + f(\mathbf{r}_{2b},\mathbf{r}_{2b}) + \bar{f} \right\}. \\
& + (1 + \mathscr{P}_{ab})2P_a^{(0)}(\mathbf{r}_1,\mathbf{r}_2)\left\{ f(\mathbf{r}_{1a},\mathbf{r}_{1a}) + f(\mathbf{r}_{2a},\mathbf{r}_{2a}) + f(\mathbf{r}_{1a},\mathbf{r}_{2a}) + f(\mathbf{r}_{2a},\mathbf{r}_{1a}) \right\} \Big\}
\end{aligned}
\quad (24)
$$

As the pair densities integrate to the number of electron pairs (here six), the normalization factor is $\mathcal{N} = \left(1 + N_a N_b \bar{f}\right)^{-1} = \left(1 + 4\bar{f}\right)^{-1}$.

Thus, we see that a substantial part of second order correlation effects on the RDMs are accounted for through functions $f$.

Now, in order to derive some independent pair approximation to Eq. (24), we replace functions $f$ by their average value $\bar{f}$, thereby reducing the expression of the pair density to

$$\mathcal{N}\left\{P^{(0)}(\mathbf{r}_1,\mathbf{r}_2)+\tfrac{1}{2}(1+\mathcal{P}_{ab})\left\{\rho_a^{(0)}(\mathbf{r}_{1a})\rho_b^{(0)}(\mathbf{r}_{2b})\left[2\omega(\mathbf{r}_{1a},\mathbf{r}_{2b})+\omega(\mathbf{r}_{1a},\mathbf{r}_{2b})^2+3\bar{f}\right]+4P_a^{(0)}(\mathbf{r}_{a1},\mathbf{r}_{2a})\bar{f}\right\}\right\}$$
$$\equiv P_{IPA}(\mathbf{r}_1,\mathbf{r}_2),\tag{25}$$

which is correct to first order only in correlation, compared with Eq. (24). The above pair density $P_{IPA}$ can equivalently be recovered by replacing each $a(\mathbf{r}_1)b(\mathbf{r}_2)$-like geminal in the uncorrelated 2-RDM by a correlated geminal $a(\mathbf{r}_1)b(\mathbf{r}_2)\left[1+\omega(\mathbf{r}_{1a},\mathbf{r}_{2b})+3\bar{f}\right]$, provided that terms beyond $\omega^2$ or $f$ are ignored in the resulting pair density. Expression (25) can thus be referred to as an independent pair approximation,[30] corresponding to the following expression for the spinless 1-RDM

$$\begin{aligned}&\rho_{1,IPA}(\mathbf{r}_1,\mathbf{r}_{1'})\\&=\mathcal{N}\left\{\rho^{(0)}(\mathbf{r}_1,\mathbf{r}_{1'})+2\rho_a^{(0)}(\mathbf{r}_{1a},\mathbf{r}_{1a'})\left[\tfrac{1}{3}f(\mathbf{r}_{1a},\mathbf{r}_{1a'})+\tfrac{5}{3}\bar{f}\right]+2\rho_b^{(0)}(\mathbf{r}_{1b},\mathbf{r}_{1b'})\left[\tfrac{1}{3}f(\mathbf{r}_{1b},\mathbf{r}_{1b'})+\tfrac{5}{3}\bar{f}\right]\right\}\end{aligned}\tag{26}$$

Comparing (26) to (23) shows that the IPA clearly underestimates dispersive correlation effects on the 1-RDM. Similar conclusions were drawn, when using a CS-like approach, see Eq. (8).

Next, we minimize the quantity

$$\langle H\rangle=\int\left[-\tfrac{1}{2}\nabla_1^2\rho(\mathbf{r}_1,\mathbf{r}_1')\right]_{r_1=r_1'}dr_1-\int\left(\frac{Z_a}{r_{1a}}+\frac{Z_b}{r_{1b}}\right)\rho(\mathbf{r}_1)dr_1+\int\frac{1}{r_{12}}P(\mathbf{r}_1,\mathbf{r}_2)\,dr_1dr_2+\frac{Z_aZ_b}{R_{ab}}\tag{27}$$

where the relevant densities have been defined in Eqs. (23) and (24) (or (25) and (26) within IPA). The orbitals $a$ and $b$ are here approximated as a single $1s$ Slater orbital. All integrals are calculated exactly, except some 2-center integrals involving VDW correction terms, which are evaluated by expanding 2-centers potential terms as successive powers of $1/R_{ab}$.

The (He)$_2$ energy is optimized with respect to variational parameters $\Delta$ and $\varsigma$, which are the average excitation energy and the $1s$ Slater orbital exponent, respectively. All results reported hereafter are in atomic units (a.u.).

Corrective terms to the potential energy turn out to be less than $10^{-8}$. The main changes in energy come from the decrease of interelectronic repulsion, which is partly balanced by an increase of kinetic energy. The obtained interaction energy is shown in Fig. 1 and further compared to those obtained within IE and IPA models. The minimization led to an interaction energy $\Delta E \approx 1.82\,10^{-5}$, together with an equilibrium position $R_{eq} = 5.68$. Optimal values of variational parameters were found to be $\varsigma = 1.68749$ (to be compared with $\varsigma_0 = 1.6875$ for isolated atoms) and $\Delta = -2.98829$. The Virial ratio at equilibrium position, e.g. $-\langle V \rangle / \langle T \rangle$, is 2 [+3 $10^{-10}$], indicating a satisfactory compensation of potential energy by the kinetic term. Results obtained for various distances are reported in table 1. Interestingly, the optimal value of $\Delta$ remains almost constant for the whole range of internuclear distances of table 1: the average value found is - 2.98828 with a standard deviation of 5 $10^{-5}$.

Note that minimizing the energy while imposing $\varsigma = \varsigma_0 = 1.6875$ yields values quasi-identical to the relaxed case as regards $\Delta E$, $R_{eq}$ and $\Delta$. However, the quantity $-\langle V \rangle / \langle T \rangle$ obtained in this case is 1.999991, denoting a slightly overestimated kinetic energy. Thus, a direct formulation of correlated density-matrices from uncorrelated ones (without relaxing subsystem electron densities) through correlation factors is questionable, given the consequences on the individual components of the total energy.

The IPA results have been obtained using RDMs defined in (25) and (26). IPA parameters were optimized following the same procedure as before (see table 1). Again, optimized $\Delta$s can be considered as constant for the whole range of internuclear distances considered (4 to 15 a.u.), though the value obtained is in this case $\Delta \approx -1.11$. The IPA gives rise to a much shorter equilibrium distance (4.99) than the former, together with a well-depth of $\Delta E \approx 1\,10^{-4}$, which is more than five times the depth found from the dispersive model of Eqs. (23), (24) and almost three times the estimated full configuration interaction value of van Mourik and Dunning [31]. The quantity $-\langle V \rangle / \langle T \rangle$ at equilibrium is now 2 [+2.5 $10^{-6}$]. These results show that IPA overestimates attractive terms and/or underestimates the kinetic energy.

Interestingly, the IPA interaction energy can however closely reproduce the results obtained from the correlated model of Eqs. (23) and (24) through an appropriate parameterization (with $\Delta$ = - 5.26, see figure 1).

Besides, accurate calculations[31,32] have shown that the equilibrium distance should be close to 5.60, with a corresponding cohesive energy of 3.48 $10^{-5}$. In comparison, an MP2 calculation within aug-ccpV5Z basis set results in 5.82 and 2.07 $10^{-5}$, respectively. This comparison shows that although dispersion forces are underestimated in our model, they lead to a quite accurate equilibrium distance (5.68), together with a reasonable cohesive energy ($\Delta E \approx 1.82\,10^{-5}$). In fact, the latter approximately corresponds to the sum of exchange and dispersion energy components $E_{exch}^{(10)} + E_{disp}^{(20)} \approx 1.89\,10^{-5}$ in the SAPT decomposition of the $(He)_2$ interaction energy.[33] We note that our result probably benefits from compensation occurring in the exact energy components. Concerning the equilibrium distance, the satisfactory agreement obtained can further be understood as repulsive energy varies much faster than the attraction energy near the equilibrium distance (typically $R_{ab}^{-12}$ vs. $R_{ab}^{-6}$), making the equilibrium distance quite independent from the attractive term. Thus, one understands how important it is to accurately describe exchange, given its consequences on the repulsive energy.

Now, in most LDA approaches, exchange is not accurately described in the midpoint region. As to illustrate this, we may substitute the IE exchange energy component contributed by $P_{IE}^{(0)}$ in the correlated pair density of Eq. (24) with a local density exchange functional $K = C \int \rho(\mathbf{r})^{4/3} d\mathbf{r}$, using various values for C, namely C = 0.7937 or 0.7386, see pp. 121 – 123 of [[34]. Minimizing the energy with respect to parameters $\Delta$ and $\varsigma$, the cohesive energies obtained are $-1.2 \ 10^{-3}$ and $-3.7 \ 10^{-3}$ and the equilibrium distances are 4.91 and 5.12, respectively. Such values are in line with those actually obtained from conventional DFT approaches (see for example the results listed in table 1 of Ref. 1]) but are in poor agreement with the values obtained using the exact IE exchange energy.

We may assume exchange effects to be reasonably accounted for in the dispersive model of Eqs. (20), (21). In this respect, the IPA approximation of Eqs. (25) and (26) benefits from the same determinantal exchange functional. Since IPA widely overestimates the interaction energy and underestimates the equilibrium distance, it is to be concluded that dispersive correlation effects must also be carefully implemented.

## V. Conclusions

A pairwise correlated wavefunction was used for deriving exact closed-expressions for the correlated density matrices. A dispersive model for the density matrices has further been proposed, correct to second order in the correlation function $\omega$. Such expressions illustrate $N$-electron effects (yet induced by a pairwise wavefunction) on the reduced densities. Application to the $(He)_2$ dimer leads to a fair cohesive energy and a quite accurate equilibrium distance. Furthermore, we put emphasis on the shortcomings arising from an independent pair approximation (which is somehow similar to the Colle-Salvetti approach): correlation corrections to the 1-electron densities are markedly underestimated, as reported elsewhere.[26] The independent pair approximation was accordingly found to overestimate (dispersive) correlation and, in turn, underestimate the equilibrium distance of $(He)_2$ when variationaly optimizing the energy. However, an appropriate parameterization of the independent pair approximation model enables to closely reproduce the cohesive energy obtained from the first dispersive model. Similarly, we can infer that an empirical parameterization of the correlation functional may artificially include $N$-electron ($N > 2$) effects as well as the neglected kinetic energy of correlation. Interestingly, such a result may partly explain the accuracy of the Colle-Salvetti's model. All the more, such a correlation functional may possibly remain accurate even far from the equilibrium geometry, as in the present case. Calculations could thus be tested for large Van der Waals systems, using an approximate correlation functional, conveniently parameterized.

Finally, we have evaluated the impact of dispersive effects, using Eq. (20), for both the $(He)_2$ and the $(CH_4)_2$ system. Dispersion effects result in additional lobes in the charge density, along the molecular axis, while slightly shifting the momentum density $n(p)$ towards higher momenta, consistently with the virial theorem. The maximum relative deviations for electron densities of $(He)_2$ were found to be $\Delta\rho/\rho \approx 6\,10^{-6}$ and $\Delta n/n \approx 5\,10^{-5}$. These small deformations can be explained by the weak polarizabilities of Helium atoms. More generally, closed-shell systems are weakly polarizable so that dispersive effects should not lead to significant corrections on 1-electron densities. In addition, applying Eq. (27) to $(CH_4)_2$ results in substantially larger maximal relative deformations. The average excitation energy was approximated using either the ionization energy or the smallest allowed excitation energy. However, the absolute deformation magnitude in position space remains similar to the $(He)_2$ case, $i.e.$ less than $1.0\,10^{-4}$ Bohr$^{-3}$ or $1.0\,10^{-3}$ Å$^{-3}$. Such magnitudes remain substantially below the current resolution of X-ray experiments ($\approx 0.05$ Å$^{-3}$). All the more, such deformations are very small compared with deformations due to intra-atomic correlation, which may contribute up to a few percent to 1-electron densities.[12,14,35] Accordingly, it is unlikely that the sole dispersive effects be observable through x-ray diffraction or Compton scattering experiments.

<u>TABLE CAPTIONS</u>

Table 1: Comparison of model interaction energies at various internuclear distances (in atomic units).

Table 1, S. Ragot, Journal of Chemical Physics

| Inter atomic distance $E - E\infty \ (\times 10^5)$ | 4.00 | 4.50 | 4.99 | 5.60 | 5.68 | 7.50 | 9.00 | 15.00 |
|---|---|---|---|---|---|---|---|---|
| Dispersive model of Eqs. (23) and (24) | 74.7336 | 15.1628 | 1.00757 | -1.79911 | -1.81549 | -0.54976 | -0.18628 | -0.00869 |
| ς (optimized) | 1.68787 | 1.68762 | 1.68752 | 1.68749 | 1.68749 | 1.68750 | 1.68750 | 1.68750 |
| IPA model of Eqs. (25) and (26) | 34.9931 | -4.4667 | -9.60614 | -7.0979 | -6.68231 | -1.46957 | -0.49448 | -0.02308 |
| ς (optimized) | 1.68757 | 1.68747 | 1.68744 | 1.68745 | 1.68746 | 1.68749 | 1.68750 | 1.68750 |
| Full CI[31] | 18.5949 | - | - | -3.48117 | - | -0.973397 | -0.31359 | -0.01330 |

Figure 1 : Model interaction energies of (He)$_2$. Dashed: Independent electron model, Eq. (15). Grey: Dispersive model of Eqs. (23) and (24). Dotted: parameterized IPA model of Eqs. (25) and (26). Dotted-dashed: variationaly optimized IPA model of Eqs. (25) and (26).

Figure 1, S. Ragot, Journal of Chemical Physics

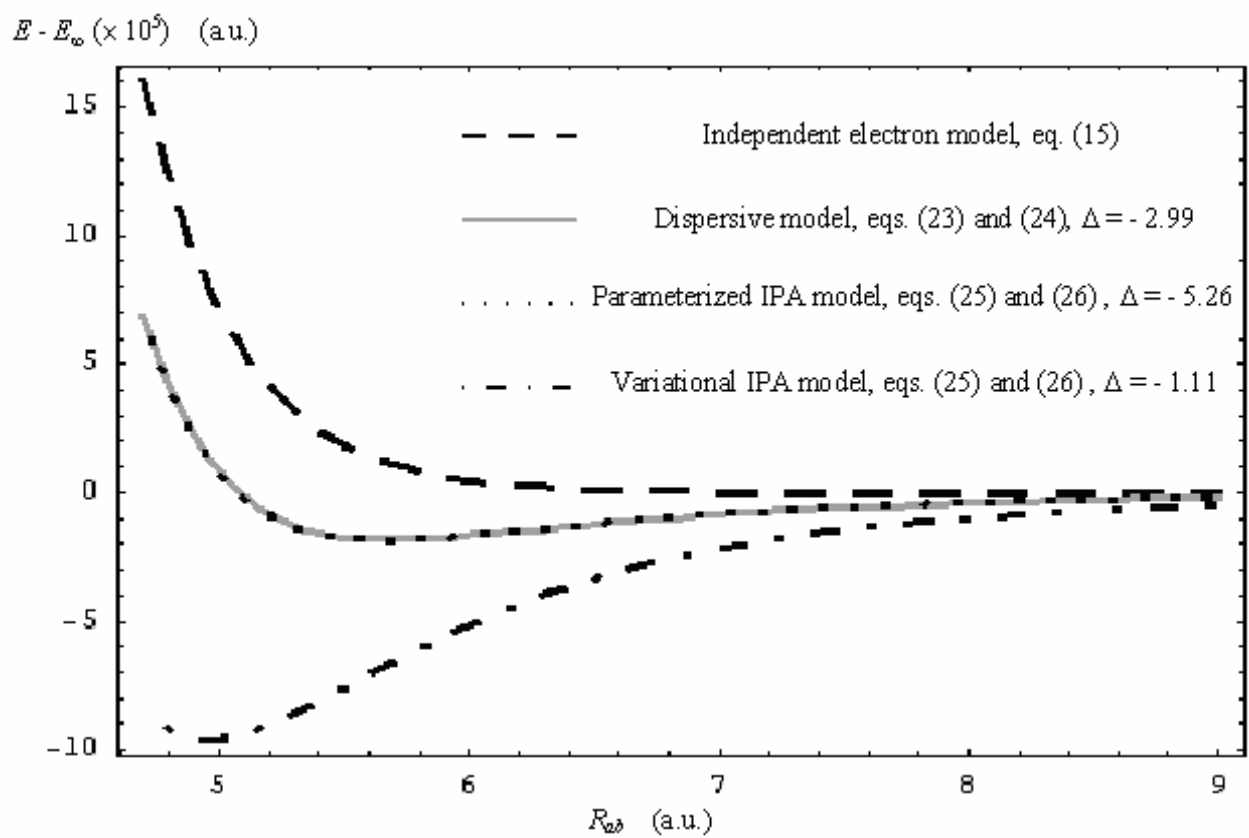